\documentclass[referee]{raa}            

\usepackage{graphicx,times}             
\usepackage{natbib}
\usepackage{tablefootnote}
\usepackage{amssymb,amsmath}
\usepackage{ulem}    
\usepackage{multirow}         
\bibpunct{(}{)}{;}{a}{}{,}

\usepackage[pagebackref=true]{hyperref}

\begin{document}

Inter-Correlation Between Sunspot Oscillations and Their Internal Structures  \title{
}

   \volnopage{Vol.0 (20xx) No.0, 000--000}      
   \setcounter{page}{1}          

   \author{Libo Fu          
      \inst{1}
   \and Zizhan Zhu
      \inst{1}
   \and Ding Yuan
      \inst{1}\thanks{Corresponding author}
   \and Jiaoyang Wang          
      \inst{1}
   \and Song Feng
      \inst{2}
   \and Sergey Anfinogentov
      \inst{3}
   }

   \institute{Institute of Space Science and Applied Technology, Harbin Institute of Technology, 
               Shenzhen 518055, China; {\it yuanding@hit.edu.cn}\\
        \and
             Faculty of Information Engineering and Automation, Kunming University of Science and Technology, Kunming 650500, China\\
        \and
            Institute of Solar-Terrestrial Physics SB RAS, Lermontov st. 126a, 664033, Irkutsk, Russia\\  
\vs\no
   {\small Received 20xx month day; accepted 20xx month day}}

\abstract{ Three- and five-minute oscillations are commonly found in any sunspot. As they are modulated by the internal thermal and magnetic structures of a sunspot, therehence, they could be used as an effective tool for sunspot seismology. 
In this paper, we investigate the properties of oscillations in sunspot groups with varying size and magnetic field, and aim to establish the relationships between sunspot oscillations and its internal structure comparatively. We selected three groups of unipolar sunspot with approximately axial-symmetric magnetic field and calculated their Fourier spectra based on the Ultraviolet(UV)/Extreme ultraviolet(EUV) emission intensity variations recorded by the Solar Dynamics Observatory/Atmospheric Imaging Assembly (SDO/AIA). 
We found that the distribution of three minute oscillation is defined by the joint effect of diverging magnetic field and the stratification of sunspot atmosphere. Its distribution could be modified by any invading magnetic structures in the umbra. Whereas the five minute oscillations are more prominent in small spots, it implies that five minute oscillation is very closely connected with umbral dynamics. 
}

 \authorrunning{Libo Fu, Zizhan Zhu \& Ding Yuan }         
 \titlerunning{Inter-Correlation Between Sunspot Oscillations and Their Structures}  

   \maketitle

%
%
\section{Introduction}           
\label{sect:intro}                    

Waves and oscillations are commonly found in sunspots and could reveal internal structure of sunspots (\citealt{khomenko2015oscillations,yuan2016stochastic}). The spatial distribution of oscillatory signal could be an effective probe of the magnetic field of a sunspot (\citealt{khomenko2006numerical,yuan2014oscillations}). Sunspot oscillations are classified into three-minute umbral wave and five-minute running penumbral waves at multiple layers of the sunspot atmosphere (\citealt{bogdan2006observational,yuan2014multi}), they are both considered as slow-mode magnetoacoustic waves propagating along the magnetic field (\citealt{de2002detection,sych2014wave}). As they perturb the plasma density (or pressure), they are commonly detected as emission intensity variations at various UV/EUV bandpasses (\citealt{yuan2016abnormal,botha2011chromospheric}). Therehence, sunspot oscillations could be used to diagnose the temperature and magnetic field structure of a sunspot.

Although sunspot oscillations were traditionally studied case by case, no statistical and comparative study was done. In this study, we compare the characteristics of sunspot oscillations with different thermodynamic structures  and magnetic field, and aim to establish the empirical correlation between the internal structure of a sunspot and the properties of the oscillatory signal. The thermodynamic structure of a sunspot describe the spatial distribution of the macroscopic parameters of sunspot plasma, such as density, temperature, pressure, ionization level, thermal conduction, viscosity etc. This would enable  the possibility of using sunspot oscillations to diagnose the internal structure of sunspots. 

This paper is organized as follows. Section 2 introduces the instrumentation and data analysis used in this study; Section 3 discusses the spatial distributions of properties of sunspots oscillation, and Section 4 presents the conclusion and discussion.


\section{Observation and Method}
\label{sect:Obs} 
\subsection{Data Selection and Calibration}

   \begin{table}
   \begin{center}
   \caption[]{ Metrics of selected sunspots}\label{Tab1}
   

    \begin{tabular}{cccccc}
     \hline\noalign{\smallskip}
   NOAA &  Location      & Start time (UT)  & Sunspot classification\footnotemark  & Approximate diameter & Size (S/M/L)  \\
   \hline\noalign{\smallskip}

 \multirow{2}{*}{11115} & \multirow{2}{*}{($-4'',541''$)}  & 2010-10-20 & \multirow{2}{*}{Hsx} & \multirow{2}{*}{$ 30''$}   & \multirow{2}{*}{M} \\
                        &                                 & 22:30      &                      &                              &                           \\
 \multirow{2}{*}{11312} & \multirow{2}{*}{($-2'',290''$)}  & 2011-10-10 & \multirow{2}{*}{Hsx} & \multirow{2}{*}{$ 60''$} & \multirow{2}{*}{L}    \\
                        &                                 & 18:00      &                      &                              &                           \\
 \multirow{2}{*}{11353} & \multirow{2}{*}{($-5'',100''$)}  & 2011-11-23 & \multirow{2}{*}{Hsx} & \multirow{2}{*}{$ 5''$}  & \multirow{2}{*}{S}    \\
                        &                                 & 9:00       &                      &                              &                           \\
 \multirow{2}{*}{11445} & \multirow{2}{*}{($-5'',260''$)}  & 2012-03-29  & \multirow{2}{*}{Hsx} & \multirow{2}{*}{$ 35''$} & \multirow{2}{*}{M}   \\
                        &                                 & 13:00      &                      &                              &                           \\
 \multirow{2}{*}{11528} & \multirow{2}{*}{($-5'',200''$)}  & 2012-07-28  & \multirow{2}{*}{Cso} & \multirow{2}{*}{$ 15''$} & \multirow{2}{*}{S}    \\
                        &                                 & 18:00      &                      &                              &                           \\
 \multirow{2}{*}{11537} & \multirow{2}{*}{($-5'',110''$)}  & 2012-08-06   & \multirow{2}{*}{Hsx} & \multirow{2}{*}{$ 15''$} & \multirow{2}{*}{S}    \\
                        &                                 & 17:00      &                      &                              &                           \\
 \multirow{2}{*}{11546} & \multirow{2}{*}{($-5'',150''$)}  & 2012-08-22  & \multirow{2}{*}{Hsx} & \multirow{2}{*}{$ 15''$} & \multirow{2}{*}{S}    \\
                        &                                 & 9:00       &                      &                              &                           \\
 \multirow{2}{*}{11569} & \multirow{2}{*}{($-5'',-300''$)} & 2012-09-15  & \multirow{2}{*}{Eac} & \multirow{2}{*}{$ 55''$} & \multirow{2}{*}{L}    \\
                        &                                 & 17:00      &                      &                              &                           \\
 \multirow{2}{*}{11579} & \multirow{2}{*}{($-5'',-270''$)} & 2012-09-30  & \multirow{2}{*}{Cso} & \multirow{2}{*}{$ 60''$} & \multirow{2}{*}{L}    \\
                        &                                 & 11:00      &                      &                              &                           \\
 \multirow{2}{*}{11591} & \multirow{2}{*}{($0'',40''$)}    & 2012-10-18 & \multirow{2}{*}{Dso} & \multirow{2}{*}{$ 50''$} & \multirow{2}{*}{L}    \\
                        &                                 & 9:00       &                      &                              &                           \\
 \multirow{2}{*}{11665} & \multirow{2}{*}{($-5'',290''$)}  & 2013-02-03   & \multirow{2}{*}{Hax} & \multirow{2}{*}{$ 55''$} & \multirow{2}{*}{L}    \\
                        &                                 & 20:00      &                      &                              &                           \\
 \multirow{2}{*}{11801} & \multirow{2}{*}{($-5'',240''$)}  & 2013-07-29  & \multirow{2}{*}{Hsx} & \multirow{2}{*}{$ 15''$} & \multirow{2}{*}{S}    \\
                        &                                 & 8:00       &                      &                              &                           \\
 \multirow{2}{*}{11899} & \multirow{2}{*}{($-5'',55''$)}   & 2013-11-18 & \multirow{2}{*}{Cko} & \multirow{2}{*}{$ 60''$} & \multirow{2}{*}{L}    \\
                        &                                 & 16:00      &                      &                              &                           \\
 \multirow{2}{*}{12107} & \multirow{2}{*}{($-5'',-370''$)} & 2014-07-05   & \multirow{2}{*}{Dko} & \multirow{2}{*}{$ 30''$}   & \multirow{2}{*}{M} \\
                        &                                 & 22:00      &                      &                              &                           \\
 \multirow{2}{*}{12186} & \multirow{2}{*}{($-5'',-430''$)} & 2014-10-13 & \multirow{2}{*}{Cso} & \multirow{2}{*}{$ 35''$}   & \multirow{2}{*}{M} \\
                        &                                 & 16:00      &                      &                              &                           \\
   \noalign{\smallskip}\hline
 \end{tabular}
 \end{center}
 \footnotesize{$^1$ In this paper, we use the 3-component McIntosh classification to represent the sunspot category, the three components represent modified Zurich class, penumbra of largest spot, distribution of sunspots (More details in \citealt{mcintosh1990classification}).}
 \end{table}

In this study, we selected 15 unipolar (or alpha) sunspots of various sizes for statistical comparison. The coordinates, start time of observation, classification, and size are listed in Table~\ref{Tab1}. A sunspot with diameter between 5 and 15 arcsec is classified as a small-sized spot, whereas the medium- and large-sized spots have a diameter of 30 to 40 arcsec and 50 to 60 arcsec, respectively.

For each sunspot, we used continuous observation of four UV/EUV channels of SDO/AIA: 1600 \AA{}, 1700 \AA{}, 304 \AA{}, and 171 \AA{}. These channels could reveal the dynamics in the sunspot atmosphere from the photosphere, chromosphere to the corona. The AIA 1600 \AA\ is the C IV spectral line, whose radiation mainly originates from the transition region and the upper photosphere. The AIA 1700 \AA\ is the C IV spectral line near the continuum, which originates from the temperature minimum and the photosphere. The AIA 304 \AA\ is the He II spectral line, which originates from the chromosphere and the transition region. The AIA 171 \AA{} originates from the quiet corona and the  upper transition region (\citealt{lemen2011atmospheric}). 

The 171 \AA\ and 304 \AA\ observation have a cadence of 12 seconds, whereas the 1600 \AA \ and 1700 \AA\ channels have a cadence of 24 seconds. Each dataset has a one-hour continuous observation with spatial resolution of 1.2 arcsec.

The data was calibrated with the standard routine (aia\_prep.pro) provided by the SolarSoftWare. This step removed the hot pixels in the CCD (Charge-Coupled Device), subtracted the dark currents, then the flat field was corrected, and each image was normalized by its exposure time. To track the sunspot, we followed the differential rotation of the sun, thereafter, the one-hour observational data was aligned to sub-pixel accuracy. 
In order to study the association between sunspot magnetic fields in various size of sunspots,
we have used the Vector Magnetic Field from HMI/SDO (\citealt{scherrer2012helioseismic}), which computed every 12 minutes (720 seconds).

\subsection{Periodicity Analyze and Peak Period Distribution}

\begin{figure}
   \centering
   \includegraphics[width=\textwidth, angle=0]{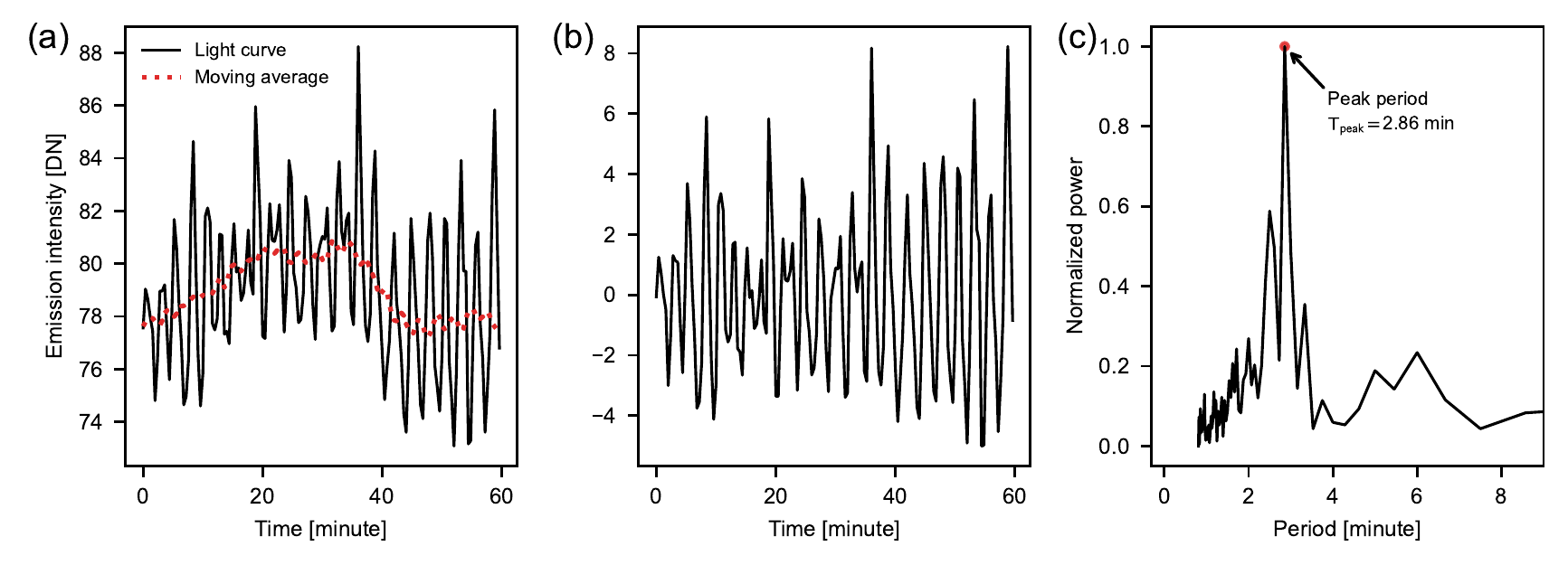}
   \caption{(a) Light curve of an example pixel in SDO/AIA 1600 \AA\ dataset and its 10-minutes moving average curve, the start time is 2011-10-10 17:59 UT. (b) De-trended light curve. (c) Normalized Fourier power spectrum of the de-trended light curve.}
   \label{Fig1}
   \end{figure}

To calculate the Fourier spectrum, we extracted the time series (or light curve) of emission intensity for each pixel. Then each time series was de-trended by removing a moving-average of a 10-minute window, and was analyzed by Fourier transform. The time series of an exemplary pixel in the SDO/AIA 1600 \AA\ and its processing procedure is illustrated in Figure~\ref{Fig1}. This analysis was applied pixel-by-pixel to each data set, then each data set was transformed into Fourier space. 

\begin{figure}
   \centering
   \includegraphics[width=\textwidth, angle=0]{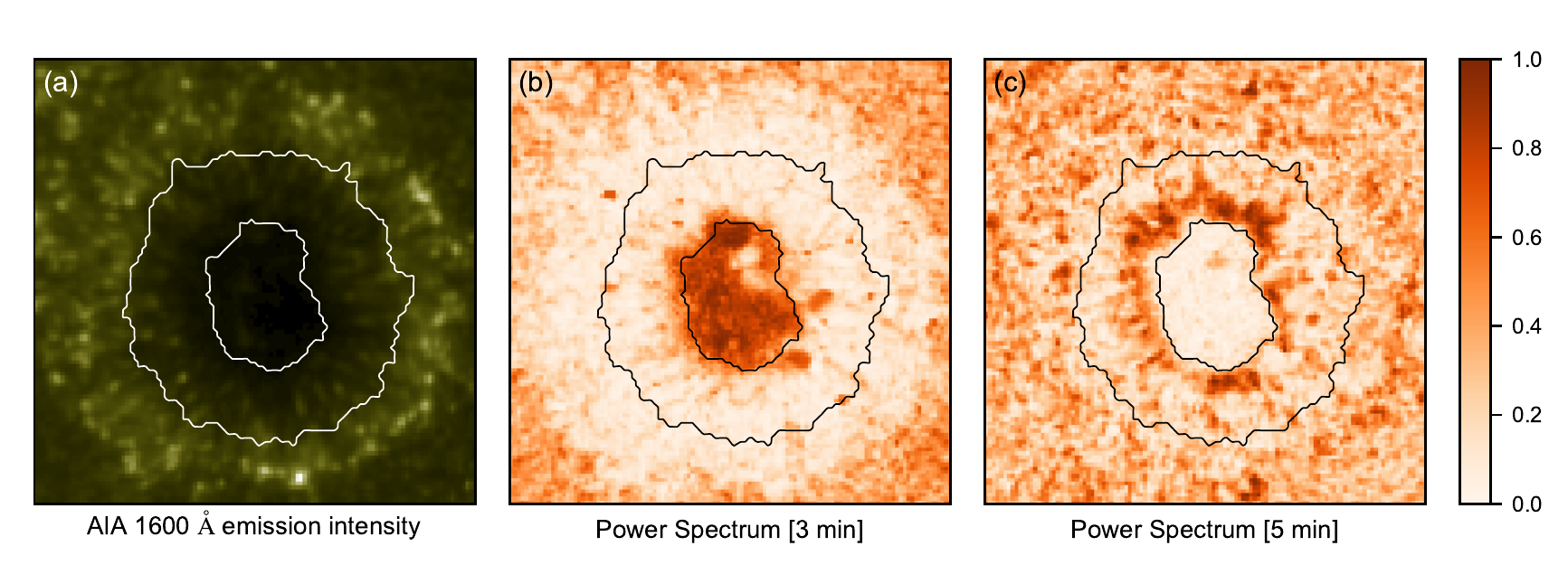}
   \caption{(a) SDO/AIA 1600 \AA{} observation of active region AR11312, recorded at 2011-10-10 17:59 UT. Two closed curves indicate the umbral and penumbral boundaries, determined with the HMI continuum intensity image.
   (b) and (c) Normalized narrowband spatial distribution averaged between 1 to 4 minutes and 4-6 minutes, denoted as $\mathrm{P_3} $ and $\mathrm{P_5} $, respectively.
   \label{Fig2}} 
\end{figure}

To calculate the power spectra for the umbral and penumbral oscillations, we calculated the average narrowband power distribution in the 3- and 5-minute range, denoted with $\mathrm{P_3}$ and $\mathrm{P_5}$, respectively. The period intervals were set to 1- to 4-minutes and 4- to 6-minutes, respectively. The average power spectrum is shown in Figure~\ref{Fig2}b and ~\ref{Fig2}c. The spatial structure of 3- and 5-minute oscillation for the 1600\AA{}, 1700\AA{}, 304\AA{} and 171\AA{} bandpass is visualized in Figure~\ref{Fig3} and \ref{Fig4}.
 

The peak period $T_{\mathrm{peak}} $ is the period corresponding to the strongest power in spectrogram. After obtaining the spectrum for each pixel point of the sunspot data separately, the period corresponding to the maximum power value in the spectrum plot is defined as the peak period for each pixel, see Figure~\ref{Fig1}.

\subsection{Profiles of Three-minute and  Five-minute Oscillation in a sunspot}

In this paper, sunspot groups of different sizes are selected in order to reduce the differences due to AIA resolution. To obtain the spatial distribution (or profile) of sunspot oscillations, we interpolated the averaged narrowband spatial distribution  and magnetic field strength over a slit across the sunspot center. The center of the sunspot umbra to the penumbra boundary in the selected image is used as a slicing slit, i.e. the circle is divided evenly into 16 parts. To reduce the error introduced by the non-circular shape of a sunspot, we took the average value over 16 slits, each slit was $22.4^{\circ}$ apart from its neighbor in the polar coordinate with origin at the sunspot center. The mean value and standard deviation of the power profiles along 16 slits was used as the profile and its uncertainty. 

To compare the power profile of sunspots with various size, we normalize the geometry of a sunspot to its radius, see Figure~\ref{Fig7} and ~\ref{Fig8}. This step was also applied to the properties of the magnetic field, see Figure~\ref{Fig6}.

\section{Results}
\label{sect:analysis}

\subsection{Power Distribution Analysis}

   \begin{figure}
   \centering
   \includegraphics[width=\textwidth, angle=0]{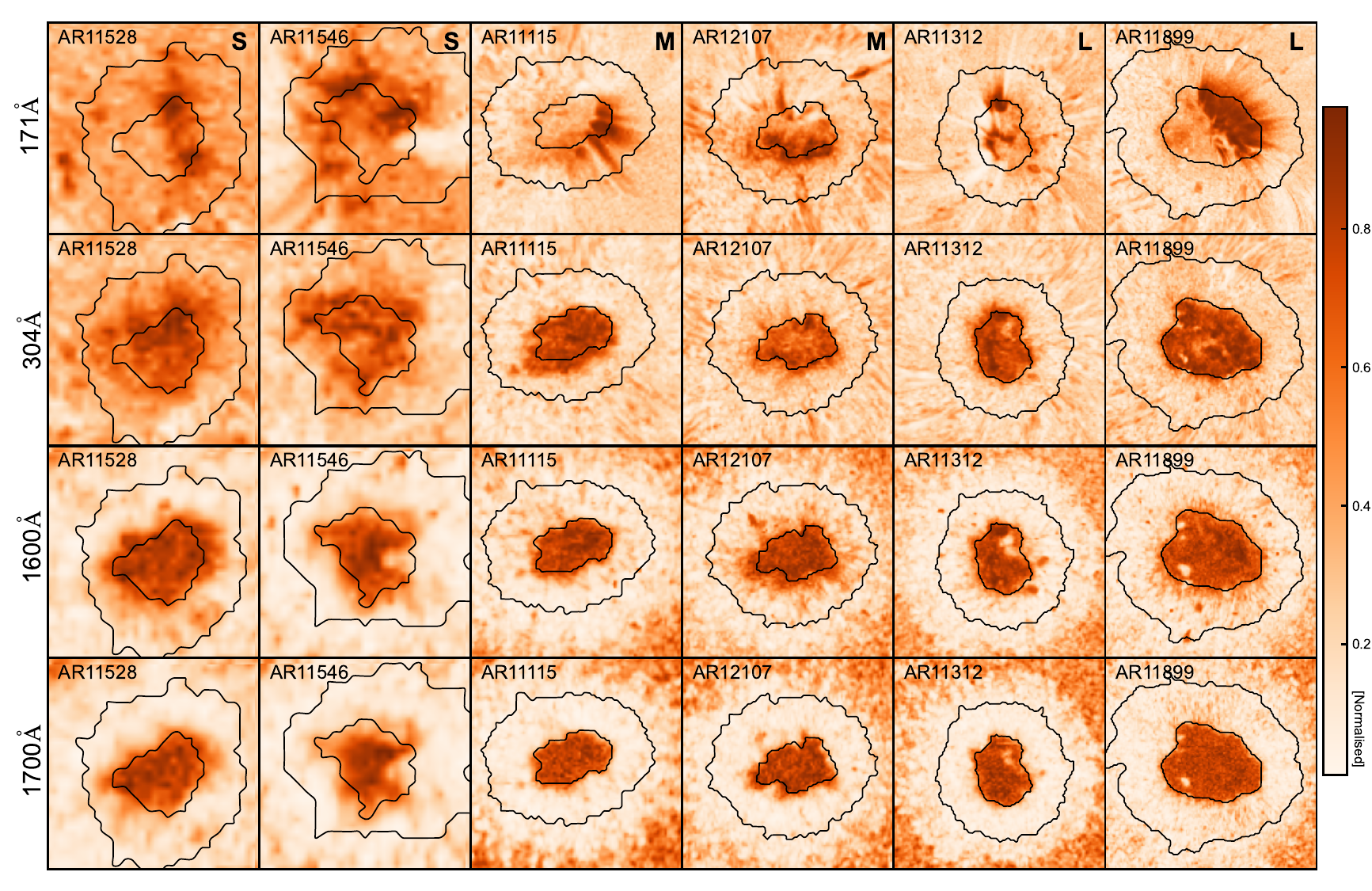}
   \caption{Spatial distribution of three-minute oscillations $ \mathrm{P_3} $ at multiple bandpasses of sunspots, observing four bands of AIA. The first two columns for small-sized sunspots, the third and fourth column for medium-sized sunspots, and the last two columns for large-sized sunspots. }
   \label{Fig3}
   \end{figure}

To visualize the performance of the observed oscillatory motion in sunspots in various bandpasses, Figure~\ref{Fig3} shows six selected sunspots with different sizes, each sunspot sample was interpolated to ensure a consistent image size for sample of sunspots. 

It could seen that in the 1600\AA, 1700\AA{}, and 304\AA{} bandpasses, the 3-minute oscillation is mainly distribution in the umbra, revealing the structure of sunspot umbra, whereas in the 171\AA{} the 3-minute oscillation follows the coronal loop structures, revealing its nature of a propagating wave. 

The 3-minute power distributions of small-sized sunspots have irregular shapes, whereas those of the medium- and large-sized spots have a more circular shape. The difference is that the large-sized sunspots reveal better fine structures in the power distribution because larger sunspots are more visible in more detail than smaller sunspots due to spatial resolution. It also could show that as the formation height of the AIA band passband increases, the area occupied by the 3-min oscillations is expanding (as shown in AR11115 and AR11312), these results are consistent with earlier results (\citealt{reznikova2012spatial})

\begin{figure}
   \centering
   \includegraphics[width=\textwidth, angle=0]{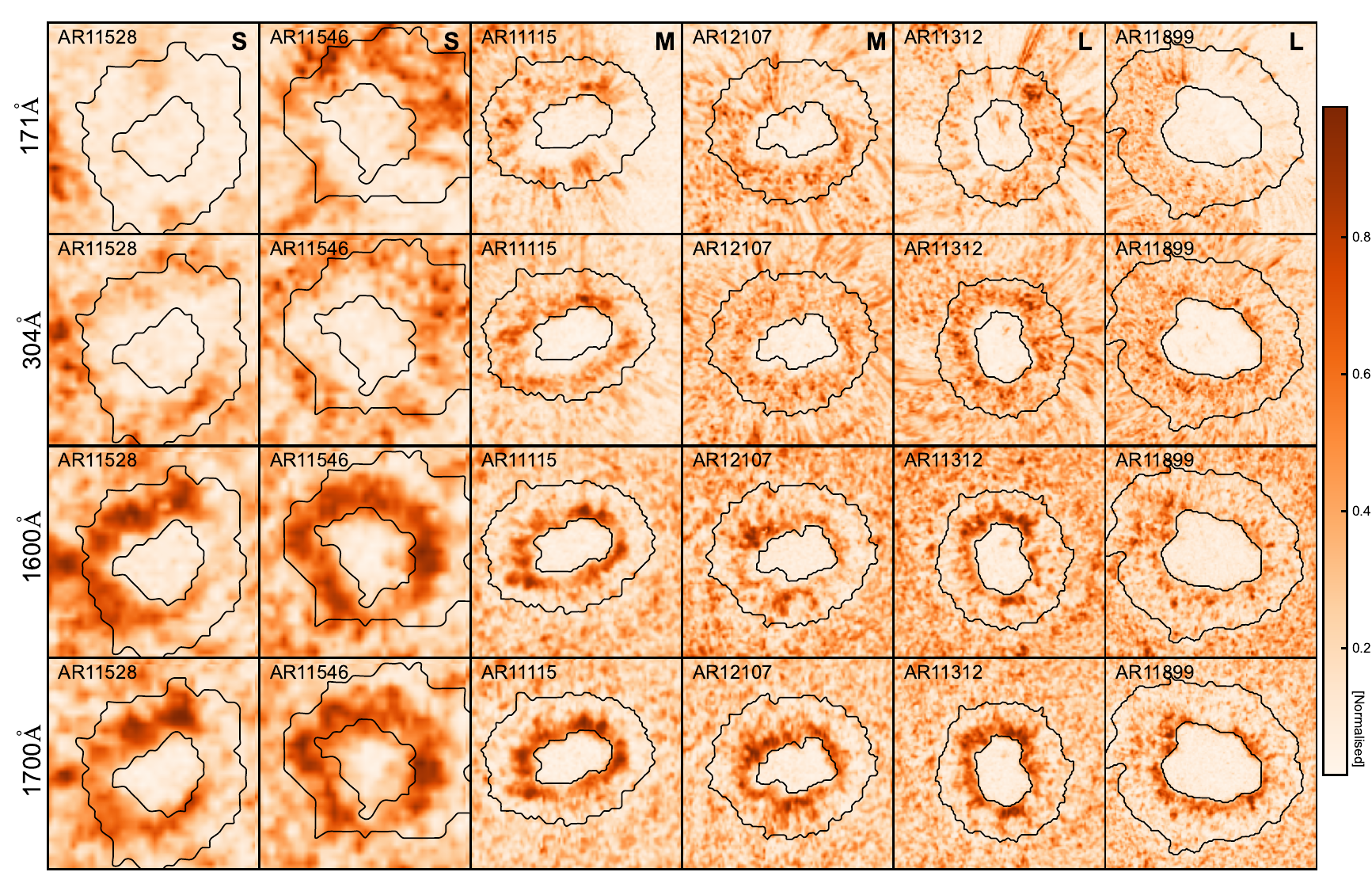}
   \caption{Spatial distribution of five-minute oscillations $\mathrm{P_5} $ at multiple band passes of sunspots. }
   \label{Fig4}
   \end{figure}

Figure~\ref{Fig4} shows the 5 min power distribution  of the sunspots, which is a 5-minute equivalent of Figure~\ref{Fig3}. The 5-minute oscillation power usually form a ring-structure around a sunspot (\citealt{kobanov2013oscillations}). In small and medium size sunspots, this structure is very apparent, whereas in large-sized sunspots, this structure become very diffuse. In some large spots (e.g AR11899), the 5-minute oscillation even disappeared.

\subsection{Distribution of Peak Period}

\begin{figure}
   \centering
   \includegraphics[width=\textwidth]{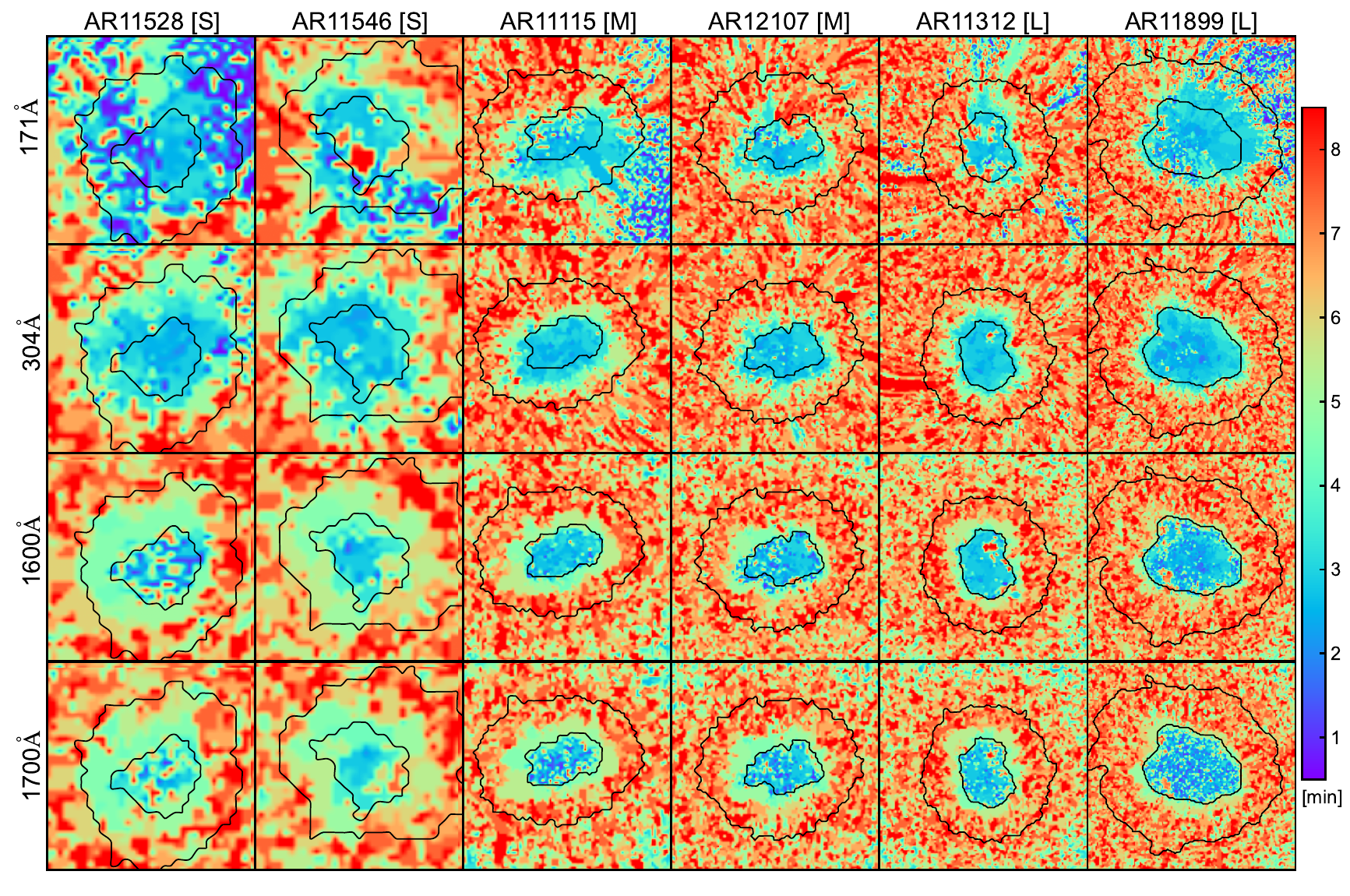}
   \caption{Spatial distribution of the peak period $T_{\mathrm{peak}}$ of the intensity oscillation in the sunspot region.}
   \label{Fig5}
\end{figure}

Figure~\ref{Fig5} presents the peak period distribution of 6 sunspots of various size and in multiple bandpasses. In a sunspot of any size, the umbral region is normally occupied by oscillations with peak period around 3-minutes, where as the penumbral region is dominated by 5-minute oscillations. The 171\AA{} data reveals a different physics, as the 5-minute oscillation could not propagate to the corona. 

In the large sunspots (e.g AR11899 and AR11312), the umbral region reveals very fine structure. The whole umbral region was filled with 3-minute oscillations and many other oscillations. Numerous dot-shaped regions were filled with either 1-minute oscillations, as studied in \citealt{wang2018high} or 5-minute oscillations. The 1-and 5-minute oscillations are uniformly distributed around the umbral region.

As the sunspot size increases, the range of high frequency regions with lower main peak period is larger in the sunspot region. Compared with the 5 minutes of $T_{\mathrm{peak}}$ spatial distribution of AIA 1600 \AA{}, the ring structure formed by the low frequency signal in the 5 minutes of $T_{\mathrm{peak}}$ spatial distribution of AIA 1700 \AA{} is more obvious, indicating that the long period signal in the penumbral region is more significant. Focusing on sunspots of different sizes at AIA 1700 \AA{}, it can be found that the ring structure is more obvious to sunspots with larger diameters. We can also find some low frequency oscillations(as in AR11312) in large and medium-sized sunspot umbra, the spatial structure can also find in the 3-min power distribution.

The 3-minute oscillation dominates in the penumbra of AIA 304 \AA{} images, unlike the mixed wave modes with 3 and 5-minute oscillations in the photosphere. In  171 \AA , intensity oscillations seem to have disappeared and the signal has been dispersed.
Normally, small size sunspots have fewer details in spatial resolution, but in fact the signal-to-noise ratio of small spots is better,  as the umbra of a spot normally darker than that of penumbra and quiet sun region. As studied in the noise analysis of SDO/AIA (\citealt{yuan2012measuring}),  the photon noise dominates in data noises, including the photon Poisson noise, compression noise, dark current noise, and despike noise, etc. The stronger emission intensity has a larger photon counts and hence a better signal-to-noise ratio, the umbra of a small spot is relatively brighter than that of a large spot, so they have a better signal-to-noise ratio.

Figure~\ref{Fig5} shows the main oscillation period of sunspots with altitude, where the main oscillations of the umbra gradually change from those with a five-minute peak period to those with a three-minute period as the atmospheric altitude increases, with oscillations in the five-minute range dominating around sunspots. Comparison with the peak period diagram of the lower photosphere (1600 \AA{}, 1700 \AA{}) proves that these oscillations are still caused by the influence of the photospheric P-mode oscillations.

 \subsection{Relationship between Sunspot Oscillations and its Internal Structure}

\begin{figure}
   \centering
   \includegraphics[width=\textwidth, angle=0]{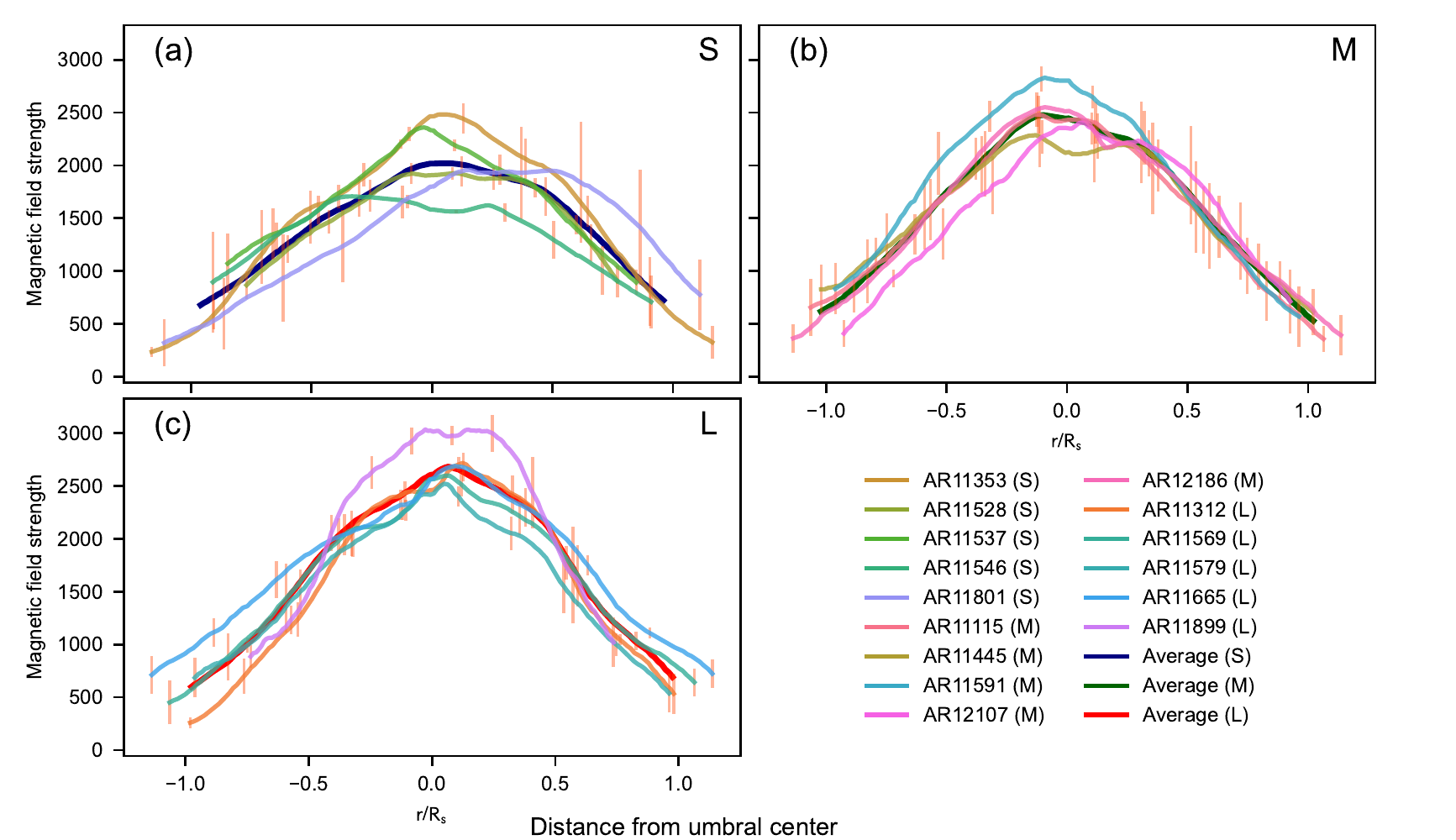}
   \caption{Magnetic field intensity slice diagram, from SDO/HMI; The horizontal coordinate measures the distance from the center of the sunspot. The value indicates the scale in units of the sunspot diameter. The vertical coordinate is the value of magnetic field strength. The the vertical bar is the standard error. }
   \label{Fig6}
   \end{figure}

\begin{figure}
   \centering
   \includegraphics[width=\textwidth, angle=0]{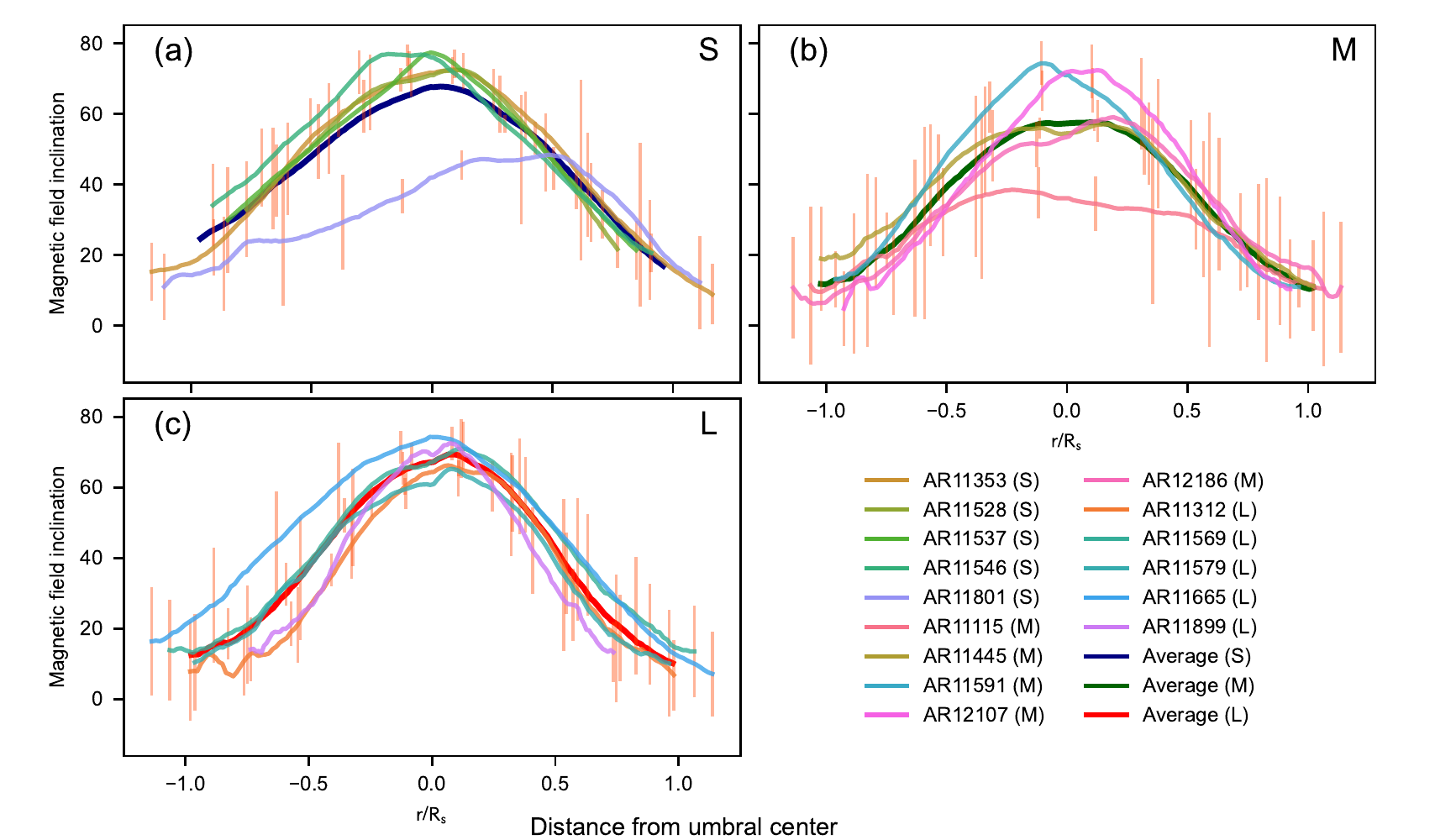}
   \caption{Magnetic field inclination slice diagram, from SDO/HMI; The vertical coordinate is inclination, limit the inclination angle to less than 90 degrees to eliminate the effect of different magnetic poles. }
   \label{Fig7}
\end{figure}

Figure~\ref{Fig6} and Figure~\ref{Fig7} shows the slice information of the magnetic field strength and inclination, from which it can be seen that the overall magnetic field distribution of sunspots shows a bell-shaped curve, the magnetic field strength of small sunspots is 500-1000 Gauss lower than that of large sunspots, and the magnetic field in the sunspot region gradually decreases about 70 degrees from the nearly vertical center of the penumbra to the edge of the penumbra.
The larger the size of the sunspot, the more compact the magnetic field structure is, the smaller the half-peak width of the bell-shaped curve is, and the smaller the standard error is. The farther the distance from the sunspot center, the larger the standard error of the slice.
\begin{figure}
   \centering
   \includegraphics[width=\textwidth, angle=0]{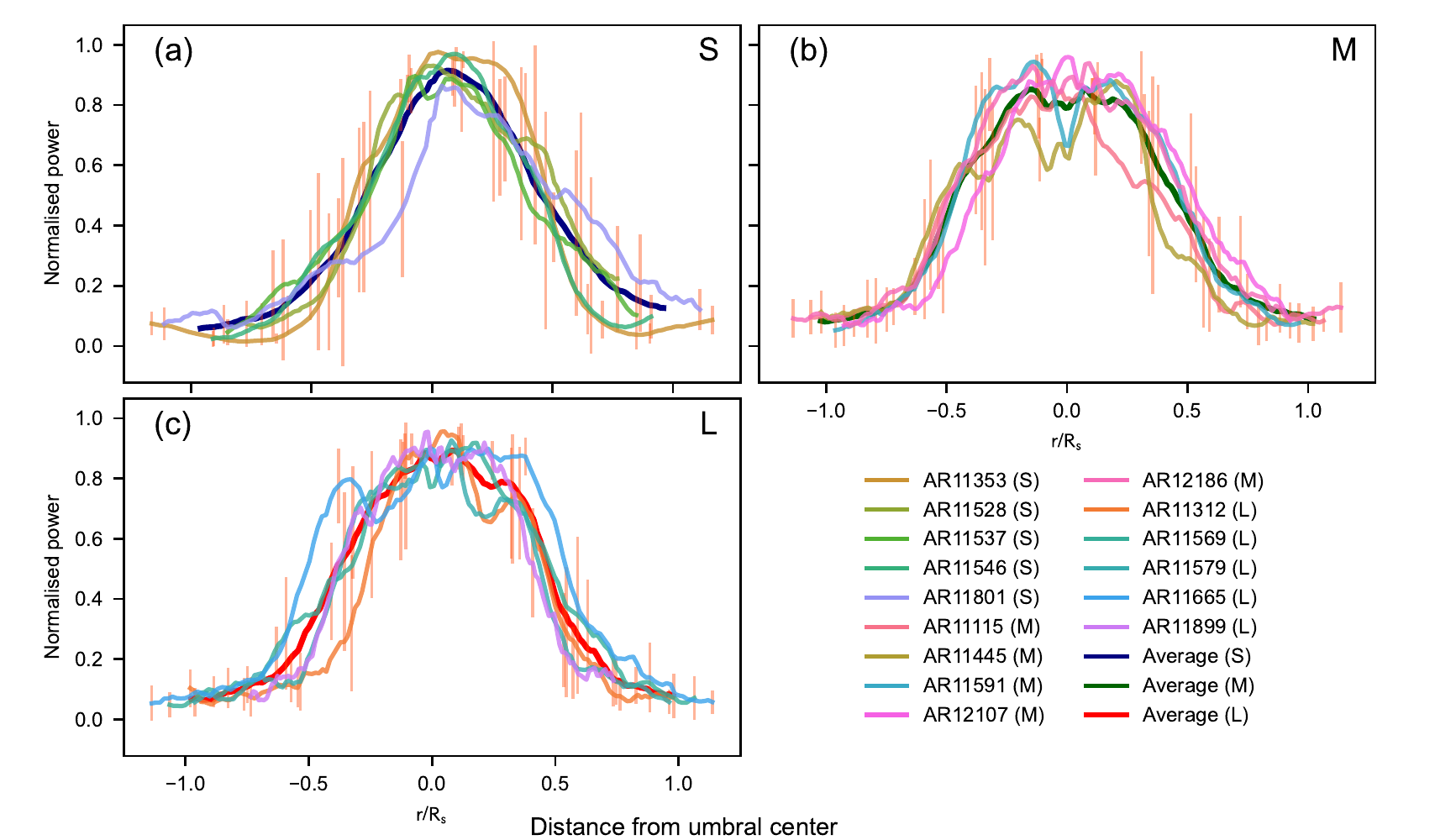}
   \caption{Three-minute narrowband power spectrum distribution, from SDO/AIA 1600 \AA. }
   \label{Fig8}
\end{figure}

\begin{figure}
   \centering
   \includegraphics[width=\textwidth, angle=0]{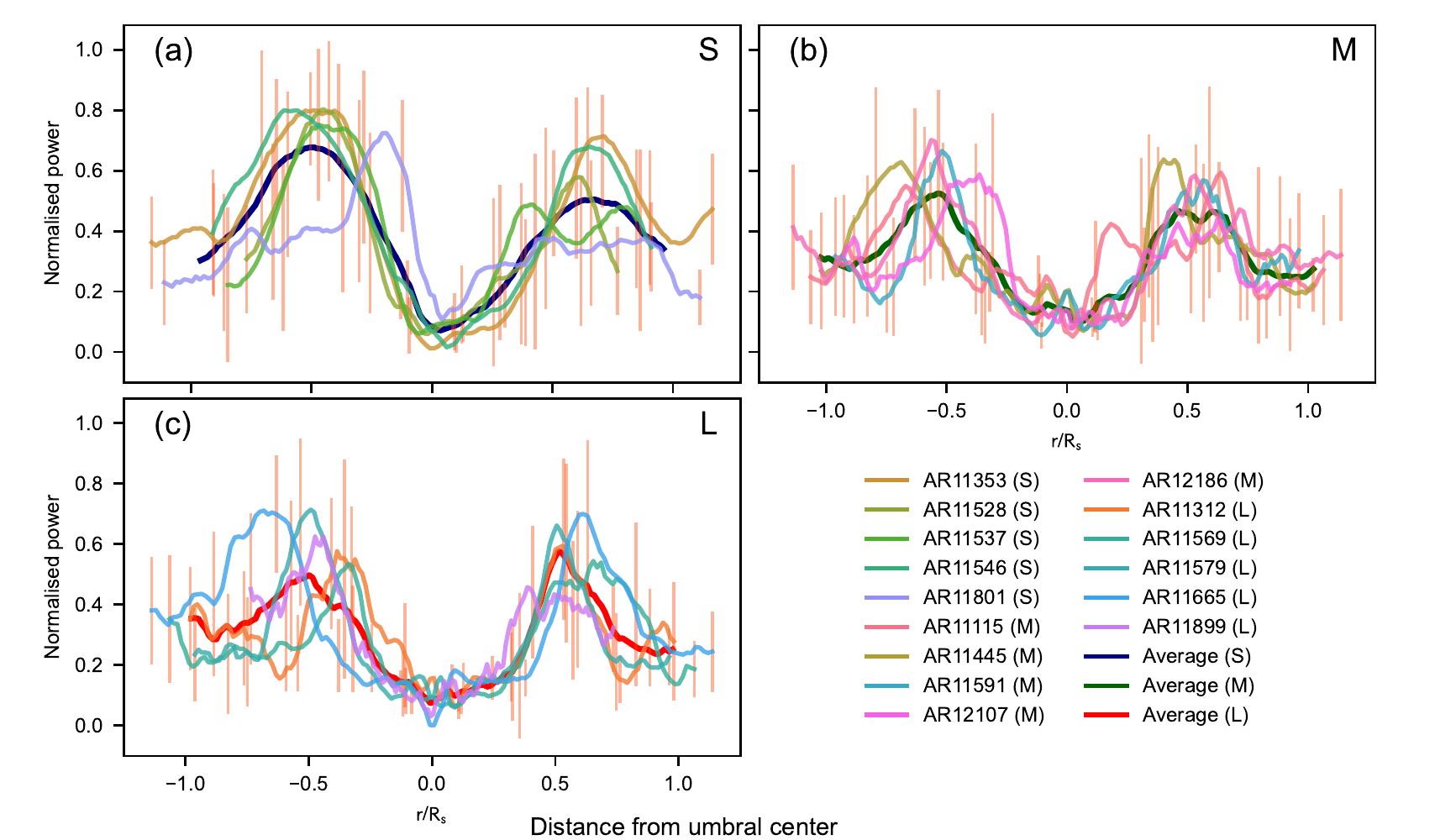}
   \caption{Five-minute power narrowband spectrum distribution, from SDO/AIA 1600 \AA.}
   \label{Fig9}
\end{figure}

Figure~\ref{Fig8} compared slices of sunspot power distribution $\mathrm{P_3}$, all sunspot curves at 1600 \AA{}, 1700 \AA{} and 304 \AA{} are similar to the bell shape. The bell-shaped diagrams are narrower for the smaller size sunspots, which have smaller half-peak widths. Overall, the further away from the sunspot center, the smaller the sunspot size and the larger the standard error of the slice.
The 171 \AA{} slice plots show no obvious pattern, and large sunspots are comparable to small sunspots in terms of error performance. 
Figure~\ref{Fig9} shows the comparison of slices of sunspot $\mathrm{P_5}$. All sunspot curves show a double-hump structure. 
Observe the different sizes of sunspot slices in Figure, small and large size sunspots have relatively insignificant structural changes, 
The amplitude of the fluctuations of medium-sized sunspot in the region of penumbra is more pronounced, this is probably due to the interaction of thermal and magnetic pressure

\section{Conclusion and Discussion}
\label{sect:conclusion}

In this study, we analyzed the oscillations of 15 sunspots of different sizes using SDO/AIA data. The aim of this paper is to analyze and compare the periodic signals of sunspot emission intensity, obtain the spatial distribution of sunspot oscillation characteristics by spectrum analysis, explore the pattern of sunspot oscillation characteristics with sunspot magnetic field and thermodynamic structure by combining the variation of sunspot magnetic field with sunspot size, and develop tools for diagnosing sunspot structure.

Our observations are as follows: (1) The 3-minute oscillations are concentrated in the umbra, the 5-minute distribution is in the penumbra. (2) The sunspot umbra 3-minute oscillations are present from the photosphere to the chromosphere, and the photosphere umbral oscillations are mixed by multiple wave modes and include 1-minute oscillations. (3) The complexity of different size of the sunspot profile is different, and the analysis of the slice curve can reveal its structure.

The power distribution analysis reveals that in the upper photosphere, chromosphere and transition region of sunspots, the power of 3 minutes oscillation is mainly concentrated in the sunspot umbra, while the power of 5 minutes oscillation is concentrated in the penumbra region, and the region with strong oscillation power shows a nearly circular or ring-like structure. Along the magnetic line direction, sunspot oscillations are transmitted in the form of slow mode magnetoacoustic waves, and the power distribution of the oscillations is basically unchanged in the solar atmosphere between the upper photosphere and the transition region; when the sunspot oscillations are transmitted to the corona, the power distribution of the oscillations starts to diverge. As shown by \citealt{reznikova2012spatial} for the magnetic field structure, the cut-off frequency of the oscillations changes due to the change of the magnetic field inclination in the sunspots, resulting in the 3 min oscillations and 5 min oscillations being confined to the sunspot umbra and penumbra. The main reason for this is that the P-mode oscillations in the photosphere interact with the strong magnetic field in the sunspot when passing through the sunspot region, and the abrupt change at the umbra-penumbra boundary is likely to be caused by the effective absorption of the p-mode oscillations.
Our conclusions are in agreement with \citealt{reznikova2012three}.

The distribution of the peak period shows that the oscillations in the chromosphere umbra are dominated by the period of 2-4 min, which can be seen in the part of the sunspot umbra to the inner umbra where the period changes more flatly until the penumbra region where the peak period increases rapidly. From the AIA 1700 \AA{} to 304 \AA{} channels, 3 minute oscillation is constantly in existence, indicates that the 3-minute oscillation originates in the photosphere  and propagates upward (\citealt{chae2017photospheric},\citealt{o2002oscillations})

The region of the umbra-penumbra boundary occupied by the 5-minute penumbra oscillation is likely a ring structure in sunspots of different sizes. This indicates that 5 minute oscillations propagate along inclined magnetic flux tube. 
In comparison with the peak period in the chromosphere, the photosphere penumbra shows a mixture of multiple dominant wave periods, and a mixture of 1 min, 3 min, and 5 min can be seen, with the 1-min peak period showing a dotted distribution in the umbra, the energy of these 1-minute oscillations may originate from deeper in the photosphere.

In AR11312 and AR11899, we can also see long-period oscillations in the umbra with a period of about 8 min, in contrast to Fig~\ref{Fig3} and ~\ref{Fig4} We can find a spatial distribution of long-period oscillations in the umbra corresponding to the absence of the power distribution in the 3-min umbral oscillation  and the distribution of power in the 5-min umbral oscillation, structures found only in mature sunspots.

The slice analysis of the sunspot magnetic field and power distribution shows that the curve variation of small sunspots is relatively smoother, and since each slice is an average of the whole sunspot slice, it can only reflect the overall variation of the sunspot. By comparing and analyzing the difference between small, medium and large sunspots, it is found that the fine variation of sunspot magnetic field can change the peak period of sunspot oscillation, and such correlation has the potential to explore the sunspot magnetic field and thermodynamic parameters.

In this research, we did the statistical distribution of 15 sunspots and found some interesting phenomena. The structure of small sunspots is likely to be too small to find more useful information because of the lack of SDO spatial resolution, and some new instruments can help us to conduct more detailed studies, e.g., Goode Solar Telescope, European Solar Telescope, . We will analyze the long-period oscillations in specific regions in sunspots afterwards, and the oscillation pattern of multiple frequency mixing in the umbra of AR12384 will be further analyzed in the next study.

\begin{acknowledgements}
This study is supported by NSFC 12173012, 12111530078 and the Shenzhen Technology project (GXWD20201230155427003-20200804151658001), S.F. is supported from the Joint Funds of the National Natural Science Foundation of China (NSFC, U1931107).
\end{acknowledgements}

\bibliographystyle{raa}

\begin{thebibliography}{20}
\providecommand\natexlab[1]{#1}
\providecommand\JournalTitle[1]{#1}

\bibitem[Bogdan \& Judge(2006)]{bogdan2006observational}
Bogdan, T., \& Judge, P. 2006, Philosophical Transactions of the Royal Society
  A: Mathematical, Physical and Engineering Sciences, 364, 313

\bibitem[Botha {et~al.}(2011)]{botha2011chromospheric}
Botha, G., Arber, T., Nakariakov, V., \& Zhugzhda, Y. 2011, The Astrophysical
  Journal, 728, 84

\bibitem[Chae {et~al.}(2017)]{chae2017photospheric}
Chae, J., Lee, J., Cho, K., {et~al.} 2017, The Astrophysical Journal, 836, 18

\bibitem[De~Moortel {et~al.}(2002)]{de2002detection}
De~Moortel, I., Ireland, J., Hood, A., \& Walsh, R. 2002, Astronomy \&
  Astrophysics, 387, L13

\bibitem[Khomenko \& Collados(2006)]{khomenko2006numerical}
Khomenko, E., \& Collados, M. 2006, The Astrophysical Journal, 653, 739

\bibitem[Khomenko \& Collados(2015)]{khomenko2015oscillations}
Khomenko, E., \& Collados, M. 2015, Living Reviews in Solar Physics, 12, 1

\bibitem[Kobanov {et~al.}(2013)]{kobanov2013oscillations}
Kobanov, N., Chelpanov, A., \& Kolobov, D.~Y. 2013, Astronomy \& Astrophysics,
  554, A146

\bibitem[Lemen {et~al.}(2011)]{lemen2011atmospheric}
Lemen, J.~R., Akin, D.~J., Boerner, P.~F., {et~al.} 2011, in The solar dynamics
  observatory (Springer), 17

\bibitem[McIntosh(1990)]{mcintosh1990classification}
McIntosh, P.~S. 1990, Solar Physics, 125, 251

\bibitem[O'shea {et~al.}(2002)]{o2002oscillations}
O'shea, E., Muglach, K., \& Fleck, B. 2002, Astronomy \& Astrophysics, 387, 642

\bibitem[Reznikova \& Shibasaki(2012)]{reznikova2012spatial}
Reznikova, V., \& Shibasaki, K. 2012, The Astrophysical Journal, 756, 35

\bibitem[Reznikova {et~al.}(2012)]{reznikova2012three}
Reznikova, V., Shibasaki, K., Sych, R., \& Nakariakov, V. 2012, The
  Astrophysical Journal, 746, 119

\bibitem[Scherrer {et~al.}(2012)]{scherrer2012helioseismic}
Scherrer, P.~H., Schou, J., Bush, R., {et~al.} 2012, Solar Physics, 275, 207

\bibitem[Sych \& Nakariakov(2014)]{sych2014wave}
Sych, R., \& Nakariakov, V. 2014, Astronomy \& Astrophysics, 569, A72

\bibitem[Wang {et~al.}(2018)]{wang2018high}
Wang, F., Deng, H., Li, B., {et~al.} 2018, The Astrophysical Journal Letters,
  856, L16

\bibitem[Yuan \& Nakariakov(2012)]{yuan2012measuring}
Yuan, D., \& Nakariakov, V. 2012, Astronomy \& Astrophysics, 543, A9

\bibitem[Yuan {et~al.}(2014{\natexlab{a}})]{yuan2014oscillations}
Yuan, D., Nakariakov, V.~M., Huang, Z., {et~al.} 2014{\natexlab{a}}, The
  Astrophysical Journal, 792, 41

\bibitem[Yuan {et~al.}(2016)]{yuan2016stochastic}
Yuan, D., Su, J., Jiao, F., \& Walsh, R.~W. 2016, The Astrophysical Journal
  Supplement Series, 224, 30

\bibitem[Yuan {et~al.}(2014{\natexlab{b}})]{yuan2014multi}
Yuan, D., Sych, R., Reznikova, V., \& Nakariakov, V. 2014{\natexlab{b}},
  Astronomy \& Astrophysics, 561, A19

\bibitem[Yuan \& Walsh(2016)]{yuan2016abnormal}
Yuan, D., \& Walsh, R.~W. 2016, Astronomy \& Astrophysics, 594, A101

\end{thebibliography}
\label{lastpage}

\end{document}